\documentclass[12pt]{iopart}
\usepackage{graphicx,bbm}
\usepackage{hyperref}
\expandafter\let\csname equation*\endcsname\relax
\expandafter\let\csname endequation*\endcsname\relax
\usepackage{amsmath}
\def\etal#1{ {\em et al.}}
\def\tit#1{}

\begin{document}
\pagenumbering{arabic}
\title{Statistics of resonances in a one-dimensional chain: a weak disorder limit}

\author{Vinayak$^{1,2}$}
\address{$^1$Instituto de Ciencias F\' isicas, Universidad Nacional Aut\' onoma de M\' exico, Cuernavaca, M\' exico}
\address{$^2$Department of Physics, Technion-Israel, Institute of Technology, Haifa 32000, Israel}

\date{\today}

\begin{abstract}
We study statistics of resonances in a one-dimensional disordered chain coupled to an outer world simulated by a perfect lead. We consider a limiting case for weak disorder and derive some results which are new in these studies. The main focus of the present study is to describe statistics of the scattered complex energies. We derive compact analytic statistical results for long chains. A comparison of these results has been found to be in good agreement with numerical simulations. 
\end{abstract}

\pacs{03.65.Yz, 03.65.Nk, 72.15.Rn}

\maketitle
\renewcommand*\thesection{\Roman{section}}

\section{Introduction}
Resonant phenomena have received much attention in atomic and nuclear physics and more recently in chaotic and disordered systems \cite{disordered, GG:00,Casati:99, KS:06,KS:08, JF:09, chaos}. Complex energies, $\tilde{{E}}_{\alpha}=E_{\alpha}-\frac{i}{2}\,\Gamma_{\alpha}$, which correspond to poles of the scattering matrix on the unphysical sheet, characterize resonances \cite{LandauL}. Resonances correspond to the long-lived quasi-stationary states which eventually decay to continuum while distribution of resonance widths, $P(\Gamma)$, determines decay of the corresponding survival probability with time.

In recent years, $P(\Gamma)$ has been a subject of investigations \cite{disordered,GG:00} for a, simple but much studied, discrete tight-binding one dimensional random chain which is coupled to a perfect lead at one side. A numerical study \cite{GG:00} shows that in a broad range of $\Gamma$, $P(\Gamma)\sim \Gamma^{-\gamma}$, where the exponent $\gamma$ is very close to $1$. Intuitively the $1/\Gamma$ behaviour can be deduced by assuming a uniform distribution for the localization centers of exponentially localized states \cite{Casati:99}. However, from analytic point of view one usually considers an infinitely long chain in which case the average density of resonances (DOR) has a well defined limit. For a finite size system, the difference between the DOR and $P(\Gamma)$ is the normalization by the system size \cite{KS:06,KS:08}. Recently, Kunz and Shapiro have derived analytic expression of the DOR for a semi-infinite disordered chain \cite{KS:08}. They have obtained an exact integral representation of the DOR which is valid for arbitrary lead-chain coupling strength. This has been further simplified for small lead-chain coupling strength where a universal scaling formula is found. In this limit they have proved the $1/\Gamma$-behavior of the DOR \cite{KS:06, KS:08}. Besides, for the continuous limit of this model an integral representation of DOR has been obtained \cite{JF:09}.

Kunz and Shapiro's work has established a universal $1/\Gamma$ law for arbitrary strength of disorder in a semi-infinite chain. Numerically one can verify $1/\Gamma$ law of the DOR, similar to what has been done by Terraneo and Guarnery \cite{GG:00} in finite samples for $P(\Gamma)$. Such verifications require the localization lengths to be much smaller than the size of the sample. In case of weak disorder an analytic result for the localization lengths is particularly useful. It comes from a second order perturbation theory. It states that the localization length is maximum near the middle of the energy band and is proportional to $W^{-2}$ where $W$ is the width of the disorder \cite{thou,krammac,KWegner:1981}. On the other hand, this result also leads to an interesting limiting situation where the localization lengths are much longer than the sample size. This is what we refer to as a weak disorder limit in this paper. This limit has scarcely been studied hitherto although it is relevant in the study of localization through resonances. Besides, there has been a believe for some sort of universality in the weak disorder limit. In this paper we address to this limit and derive analytic results which describe the statistics of resonances. Our work probes a fresh area and studies a weak disorder limit which has never been addressed before.   

For open systems, instead of studying the scattering matrix in a complex plane we follow an alternative approach where one solves the Schr\"{o}dinger equation by describing a particle ejected from the system or equivalently with a boundary condition of outgoing waves (Siegert boundary condition \cite{Siegert}). In this approach one naturally turns up to a problem of solving a non-Hermitian effective Hamiltonian which admits complex eigenvalues $\tilde{{E}}_{\alpha}$ \cite{GG:00, KS:06, KS:08, HKF:09}. For details of such non-Hermitian effective Hamiltonians, we refer to a recent study \cite{JF:10} and references therein.

We derive the statistics which describe scattered complex energies of disordered chain around those regular ones which correspond to an open chain without any disorder (clean chain). For instance, we derive average of square of the absolute values of the shifts in complex energies from the regular ones over all realizations of the set of random site energies. Similarly we obtain results for the statistics of real and imaginary parts of those shifts. These results lead to compact expressions for long chains. To show the generality of our approach we also derive these results for the so-called parametric resonances which have been particularly useful in numerical studies \cite{GG:00}. Finally, we give numerical verifications of our analytic results. 

The paper is organized as follows. Although the system and its effective Hamiltonian have been nicely explained earlier in \cite{GG:00,KS:06,KS:08}, for the sake of completeness of this paper we will describe these briefly in section II. In the same section we will also describe the exact and the parametric resonances. In Sec. III we will derive result for resonances in an open-clean chain of finite length, in terms of a polynomial equation. For long chains, we will solve this polynomial equation in the leading order of the inverse of the length. In Sec. IV we will use the perturbation theory to obtain the first and the second order corrections in the complex energies for a weak disorder. In Sec. V we will calculate statistics of the scattered complex energies. In the same section we will simplify our results for long chains and obtain compact expressions. In Sec. VI we will briefly discuss about the numerical methods to calculate complex energies of non-Hermitian effective Hamiltonians and numerically verify our analytical results. This will be followed by the conclusion in Sec. VII.

\begin{figure}
        \centering
               \includegraphics [width=0.5\textwidth]{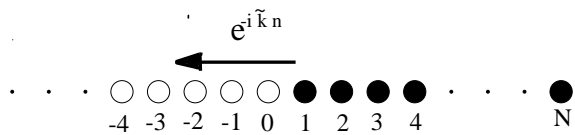}
                \caption{A one-dimensional disordered chain with $N$ sites, represented in the figure by black dots, is coupled to a lead. Open circles represent sites of the lead. The outgoing plane wave is shown by the arrow where $0<\Re\{\tilde{k}\}<\pi$ and $\Im \{\tilde{k}\}<0$, so that it propagates left in the lead and its amplitude grows in the lead.}
\label{System}
\end{figure}


\section{Model and Its Effective Hamiltonian}

A discrete tight-binding one dimensional chain of length $N$ (shown by positive integers, $n=1,\,2,...,\,N$, used for indexing the sites of the chain in Fig. \ref{System}) is connected to an outer world (represented by a perfect lead which sites are shown by a zero and negative integers, $n=0,\,-1,\,-2,...$). Each site of the chain has the site energy $\epsilon_{n}$ where $\epsilon_{n}$ are statistically independent random variables chosen from some symmetric distribution. Each nearest neighbor site of the chain as well as of the lead is coupled by a hopping amplitude $t$. The hopping amplitude for the pair $n=0$ and $n=1$ is $t'$ which takes values from $t'=0$ (closed chain) to $t'=t$ (fully coupled chain). With this hopping, a particle, which is initially located somewhere in the chain, eventually escapes to the outer world.

Now we write down the Schr\"{o}dinger equation for the entire system,
\begin{eqnarray}
\label{Sch1}
-t\psi_{n+1}-t\psi_{n-1}&=&\tilde{\mathcal{E}}\psi_{n},~~~~~~~~\text{for $n<0$},
\\
\label{Sch2}
-t\psi_{-1}-t'\psi_{1}&=&\tilde{\mathcal{E}}\psi_{0}, ~~~~~~~~\text{for $n=0$},
\\
\label{Sch3}
-t'\psi_{0}-t\psi_{2}+\epsilon_{1}\psi_{1}&=&\tilde{\mathcal{E}}\psi_{1},~~~~~~~~\text{for $n=1$},
\\
\label{Sch4}
-t\psi_{n-1}-t\psi_{n+1}+\epsilon_{n}\psi_{n}&=&\tilde{\mathcal{E}}\psi_{n},~~~~~~~~\text{for $2\leq n\leq N$.}
\end{eqnarray}
In order to avoid cluttering of notations we always represent quantities corresponding to disordered system by {\it script letters} while quantities for the clean system are represented in usual math notations. Tilde is used to discriminate the open system case from the closed one. Equation (\ref{Sch1}) is for the lead where $\epsilon_{n}=0$. Equations (\ref{Sch2}, \ref{Sch3}) describe the lead-chain coupling and Eq. (\ref{Sch4}) is for the chain. As in, \cite{KS:08} we solve Eqs. (\ref{Sch1}-\ref{Sch4}) with a boundary condition of an outgoing plane wave in the lead, i.e., $\psi_{n_{\leq 0}}\propto \exp(-i\tilde{k}n)$ where $0<\Re\{\tilde{k}\}<\pi$ and $\Im \{\tilde{k}\}<0$.  The condition on $\Re\{\tilde{k}\}$ ensures that the outgoing wave propagates to left in the lead. The condition on $\Im \{\tilde{k}\}$ is considered so that the amplitude of the resonance wave function grows in the lead. It comes from Eq. (\ref{Sch1}) that the complex energy $\tilde{\mathcal{E}}$ is related to the complex wave vector $\tilde{k}$ via the dispersion relation $\tilde{\mathcal{E}}=-2 t \cos(\tilde{k})$. Now we eliminate all $\psi_{n}$ for $n<1$ from Eqs. (\ref{Sch1}-\ref{Sch4}) and obtain
\begin{equation}\label{reduced}
-t\psi_{n+1}-t\psi_{n-1}+\tilde{\epsilon}_{n}\psi_{n}=\tilde{\mathcal{E}}\psi_{n},
\end{equation}
where 
\begin{equation}\label{energy}
\tilde{\epsilon}_{n}=\epsilon_{n}-t\eta\, \exp(i\tilde{k})\delta_{n1},
\end{equation}
for $n=1,\,2,...,\,N$. The parameter $\eta=(t'/t)^2$ measures the coupling strength to the outside world. 

An effective Hamiltonian defined by the Eq. (\ref{reduced}) is non-Hermitian. For instance, if $\mathcal{H}$ is the $N\times N$ tridiagonal Hermitian matrix which represents the Hamiltonian of the closed-disordered chain then one may write the effective Hamiltonian, $\tilde{\mathcal{H}}$, as
\begin{equation}\label{Hamil}
\tilde{\mathcal{H}}=\mathcal{H}-t\,\eta\,\lambda(\tilde{k})\, P.
\end{equation}
Here $P=|1\rangle\langle1|$ is the projection for site $n=1$ and $\lambda=\exp(i\tilde{k})$. The above non-Hermitian effective Hamiltonian has been first obtained by Terraneo and Guarnery \cite{GG:00}. The underlying result here is that the same relation (\ref{Hamil}) is valid for any Hermitian $\mathcal{H}$ representing a (closed) quantum system \cite{JF:10} which has $N$-dimensional state space. Resonances are characterized by the complex eigenvalues, $\tilde{\mathcal{E}}_{\alpha}$, of $\tilde{\mathcal{H}}$.

Note here dependency of $\tilde{\mathcal{H}}$ on the complex wave vector $\tilde{k}$ which is related to the complex energies via the dispersion relation mentioned above - this is not a standard eigenvalue problem. To standardize this problem ``parametric resonances'' are often used as an alternative. In this approach the dependence of $\lambda$ on $\tilde{k}$ is typically neglected, reducing thereby the problem of finding the eigenvalues of the effective Hamiltonian at chosen value of $\tilde{k}$. As expected, parametric resonances yield approximate statistical results which are close to those for the exact resonances in strongly localized regime \cite{GG:00}. Parametric resonances depend on a chosen parameter, for instance let $\tilde{k}=k_{0}$ and we fix it in the middle of the energy band, $k_{0}=\pi/2$. Writing explicitly
\begin{equation}
\lambda(\tilde{k})=
\begin{cases}
\exp(i \tilde{k}), & \text {for exact resonances,}
\\
i, & \text {for parametric resonances.}
\end{cases}
\end{equation}

From now on we set the energy scale by taking $t=1$, denoting the complex variable $\tilde{\mathcal{E}}/t$ by $\tilde{\mathcal{Z}}$. We denote the Hamiltonian matrix representing the closed-clean chain by $H$. It differs from $\mathcal{H}$ only at the diagonal as, for the clean chain case, all the site energies are zero. Calculation of the eigenvalues of $H$ is a standard exercise where one derives $z_{\alpha}=-2 \cos[\alpha\pi/(N+1)]$ for $\alpha=1,...,N$.

Before going into a detail treatment to the problem, we should first sketch the outline of our approach. We are interested in a weak disorder regime. Since our approach rely on perturbation theory, we need complex energies of open-clean chain, i.e., the $\tilde{z}_{\alpha}$s. So we will begin with calculating the resonances for open-clean chain of finite length. Then we will do the perturbation series expansion up to the second order of strength of the disorder. This will be followed by the derivation of the statistical results. Finally, we will consider the large-$N$ limit of these results.


\section{Open-clean Chain}
We begin with defining the resolvent $\tilde{\mathcal{G}}(z)=(z-\tilde{\mathcal{H}})^{-1}$. Using Eq. (\ref{Hamil}) we may also write
\begin{equation}\label{resolvent}
\tilde{\mathcal{G}}(z)=(z-\mathcal{H}+\eta\,\lambda\,P)^{-1}.
\end{equation}
For the open-clean chain we define the resolvent
\begin{eqnarray}
\tilde{G}(z)&=&(z-H+\eta\,\lambda\,P)^{-1}
\nonumber
\\
&=&
(1+\eta\,\lambda\,GP)^{-1}\,G,
\end{eqnarray}
where we have introduced $G(z)=(z-H)^{-1}$ as the resolvent for the ``unperturbed" closed-clean chain. Resonances correspond to the singularities of the matrix $\tilde{G}_{mn}(z)$, or to the roots of the secular equation
\begin{equation}\label{charcpol}
F(z)=0
=1+\eta\lambda G_{11}(z),
\end{equation}
where $G_{11}$ is the $\{1,\,1\}$ element of the matrix $G$ in site representation. ($G_{nm}(z)=\langle n|(z-H)^{-1}|m\rangle$.)

\begin{figure}[!t]
        \centering
               \includegraphics [width=0.75\textwidth]{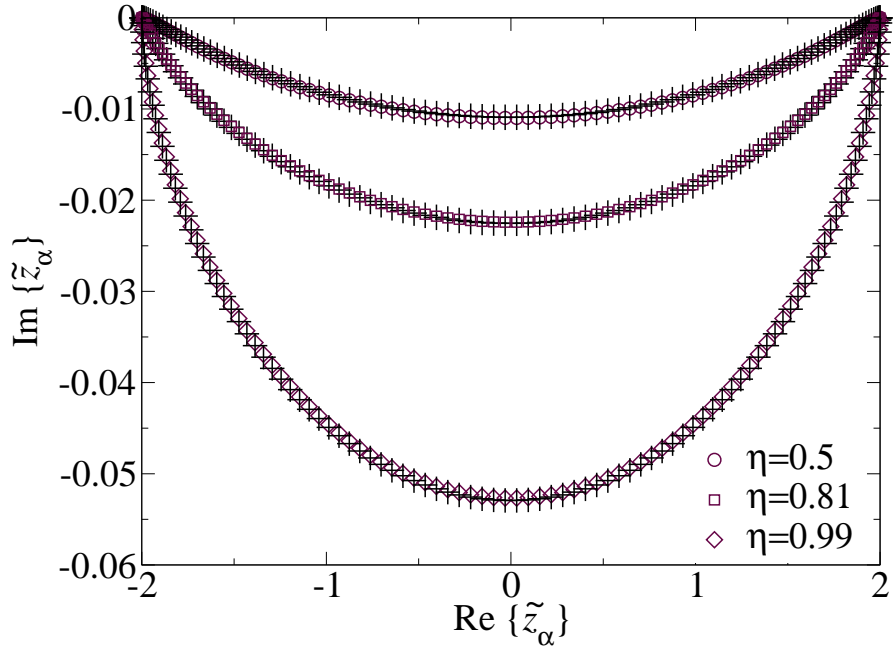}
                \caption{Comparison of the result (\ref{resz}) (pluses) with the numerical solution of the polynomial equation (\ref{fincharpol}) (circles, squares and diamonds) for the exact resonances where $\eta=0.5,\,0.81$ and $0.99$. We have considered $N=100$.}
\label{ansatz-exact}
\end{figure}

To obtain $G_{11}$ for finite $N$, we use the ordinary difference equation (ODE), 
\begin{equation}\label{ODE}
\psi_{n+1}+\psi_{n-1}+z\,\psi_{n}=0,
\end{equation}
with the boundary conditions $\psi_{0}=\psi_{N+1}=0$. This equation is obtained from Eq. (\ref{reduced}) by setting all $\epsilon_{n}=0$.  Next we consider $u_{n}(z)$ and $v_{n}(z)$ to be the two linearly independent functions which satisfy the ODE
\begin{eqnarray}\label{un}
u_{n+1}+u_{n-1}+z\,u_{n}&=&0,
\\
\label{vn}
v_{n+1}+v_{n-1}+z\,v_{n}&=&0,
\end{eqnarray}
where $u_{0}=v_{N+1}=0$. Since norm of $u_{n},\,v_{n}$ is arbitrary, we fix $u_{1}=v_{N}=1$. Further we claim that the resolvent is given by
\begin{equation}\label{G0uv}
G_{nm}=-\dfrac{u_{n}\,v_{m}\Theta(m-n)+u_{m}\,v_{n}\Theta(n-m)}{W_{n}}.  
\end{equation}
Here $\Theta(n)$ is the unit-step function and $W_{n}=u_{n}v_{n-1}-u_{n-1}v_{n}$ is the Wronskian. Using Eqs. (\ref{un}, \ref{vn}) it is straight forward to see that the Wronskian is independent of $n$. One can also check that
\begin{equation}
G_{n+1m}+G_{n-1m}+z\,G_{nm}=\delta_{nm}. 
\end{equation}
We now set $u_{n}=v_{N+1-n}$ to match the initial value problem (\ref{un}, \ref{vn}) to the boundary value problem (\ref{ODE}). We find
\begin{equation}\label{Gun}
G_{11}=-\dfrac{u_{N}}{u_{N+1}}. 
\end{equation}
The ODE (\ref{un}) is satisfied by the Chebyshev polynomial of the second kind, $U_{m}(-z/2)$, defined as
\begin{equation}
U_{m}(x)=\dfrac{\sin[(m+1)\cos^{-1}(x)]}{\sin[\cos^{-1}(x)]}, 
\end{equation}
for $U_{0}(x)=1$ and $U_{1}(x)=2x$. Since we have fixed $u_{1}=1$, therefore $u_{n}=U_{n-1}$, thus we can write Eq. (\ref{Gun}) as
\begin{equation}\label{GF}
G_{11}=-\dfrac{U_{N-1}(-z/2)}{U_{N}(-z/2)}=-\dfrac{\sin[N\,k]}{\sin[(N+1)k]}.
\end{equation}
Here the last equality follows from the energy dispersion relation. Using Eq. (\ref{GF}) in Eq. (\ref{charcpol}), we end up with an algebraic equation 
\begin{equation}\label{fincharpol}
F(z)
=0=
1-\eta\lambda\dfrac{U_{N-1}(-z/2)}{U_{N}(-z/2)}.
\end{equation}

\begin{figure}
        \centering
               \includegraphics [width=0.75\textwidth]{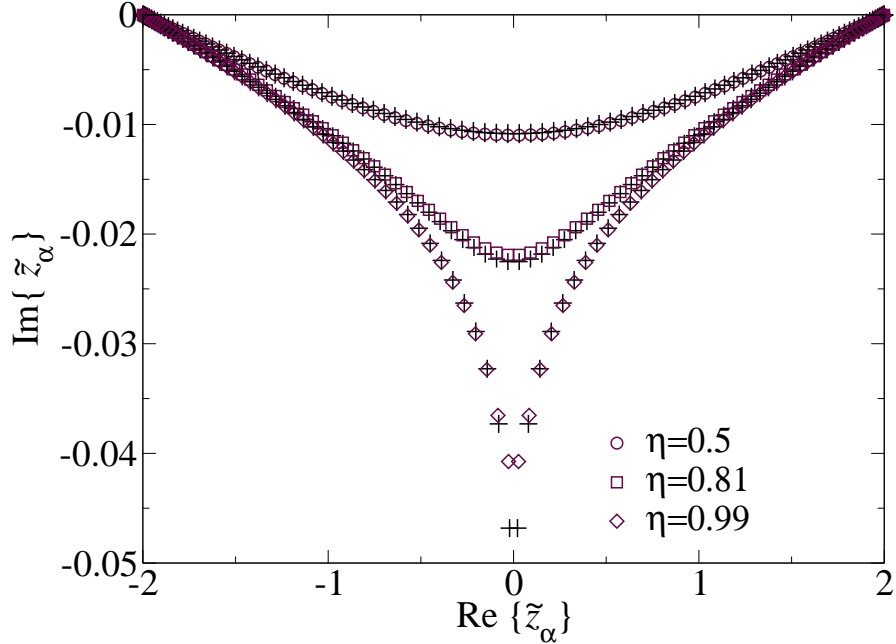}
                \caption{Repeated on the same pattern of Fig. \ref{ansatz-exact} but for parametric resonances.}
\label{ansatz-parametric}
\end{figure}

Zeros of $F(z)$ are the roots of a polynomial of order $N$. For exact resonances Eq. (\ref{fincharpol}) can be easily transformed into
\begin{equation}\label{Az}
[\mathsf{a}(z)]^{(2N+1)}=
\dfrac{\mathsf{a}(z)^{-1}-\eta\,\mathsf{a}(z)}{1-\eta},
\end{equation}
where
\begin{equation}\label{Aexp}
\mathsf{a}(z)=-\exp[ik(z)].
\end{equation}
In order to solve Eq. (\ref{Az}), we propose an ansatz assuming that opening of the system at one end causes $\mathcal{O}(N^{-1})$ complex corrections to the $k_{\alpha}$'s. Let
\begin{equation}\label{ansatz}
 \tilde{k}_{\alpha}=k_{\alpha}+\dfrac{\Phi_{\alpha}}{N},
\end{equation}
where $\Phi_{\alpha}$ is a complex quantity and $k_{\alpha}=\alpha\pi/(N+1)$. Inserting this ansatz into Eqs. (\ref{Az}, \ref{Aexp}) we obtain
\begin{equation}\label{res}
 \tilde{k}_{\alpha}= k_{\alpha}-
\dfrac{i}{2N}\,
\text{ln}\left[\Omega(k_{\alpha};\eta)\right]+\mathcal{O}\left(\dfrac{1}{N^2}\right),
\end{equation}
where 
\begin{equation}\label{Omega}
\Omega(k;\eta)=
\dfrac{1-\eta \,e^{2i\,k_{\alpha}}}{1-\eta}.
\end{equation}
Now, up to $\mathcal{O}(N^{-1})$, $\tilde{z}_{\alpha}$ may be written as
\begin{equation}\label{resz}
 \tilde{z}_{\alpha}=-2\cos(k_{\alpha})-
\dfrac{i\sin(k_{\alpha})}{N}\,
\text{ln}(\Omega).
\end{equation}
The same result can be obtained for the parametric resonances, after repeating the similar steps, but with different $\Omega$:
\begin{equation}\label{Omgpara}
\Omega(k_{\alpha};\eta)= 
\dfrac{1-i\eta \,e^{i\,k_{\alpha}}}
{1-i\eta \,e^{-i\,k_{\alpha}}}.
\end{equation}

\begin{figure}
        \centering
               \includegraphics [width=0.75\textwidth]{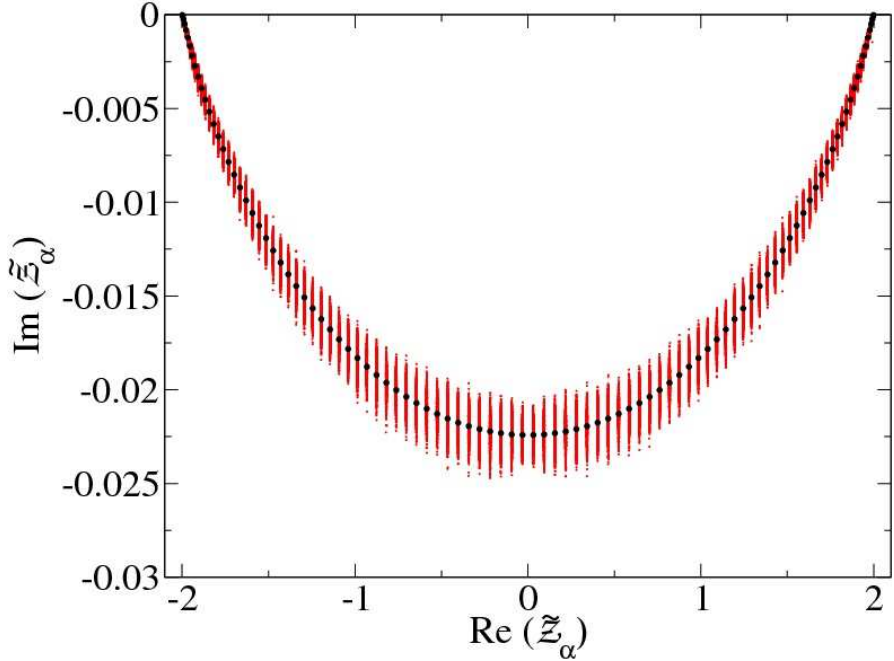}
                \caption{Scatter plot for exact resonances where $N=100$, $\eta=0.81$. Dense points in the graph represent exact resonances in the disordered chain for $2500$ realizations where $W=0.015$. These are scattered around dots which represent exact resonances in the clean chain.}
\label{scatter_exact}
\end{figure} 

One should bear in mind that there is no resonance for $\eta=1$, as the system is fully coupled to the lead. However, for parametric resonances, one artificially gets resonances even when $\eta=1$. Note that the result (\ref{resz}) is symmetric about the imaginary axis for both cases.
In Fig. \ref{ansatz-exact} and Fig. \ref{ansatz-parametric} we compare the numerical solutions of the polynomial equation (\ref{fincharpol}) with our results (\ref{resz}, \ref{Omega}, \ref{Omgpara}), for $N=100$,  $\eta=0.50,\,0.81$ and $0.99$ and $N=100$, respectively for exact and parametric resonances. Eq. (\ref{fincharpol}) has been solved by using the Newton's method with the initial guess $\tilde{k}_{\alpha}=k_{\alpha}$. These figures show that our result (\ref{resz}) is close to the numerical solution. The agreement gets better as $\eta\rightarrow1$ (not shown here separately). However, the ansatz (\ref{ansatz}) is not valid near the band edges. Moreover, the agreement fails for parametric resonances near the middle of the band as $\eta\rightarrow 1$; see Fig. \ref{ansatz-parametric} for $\eta=0.99$. 


\section{The Weak disorder limit}
\begin{figure}
        \centering
               \includegraphics [width=0.75\textwidth]{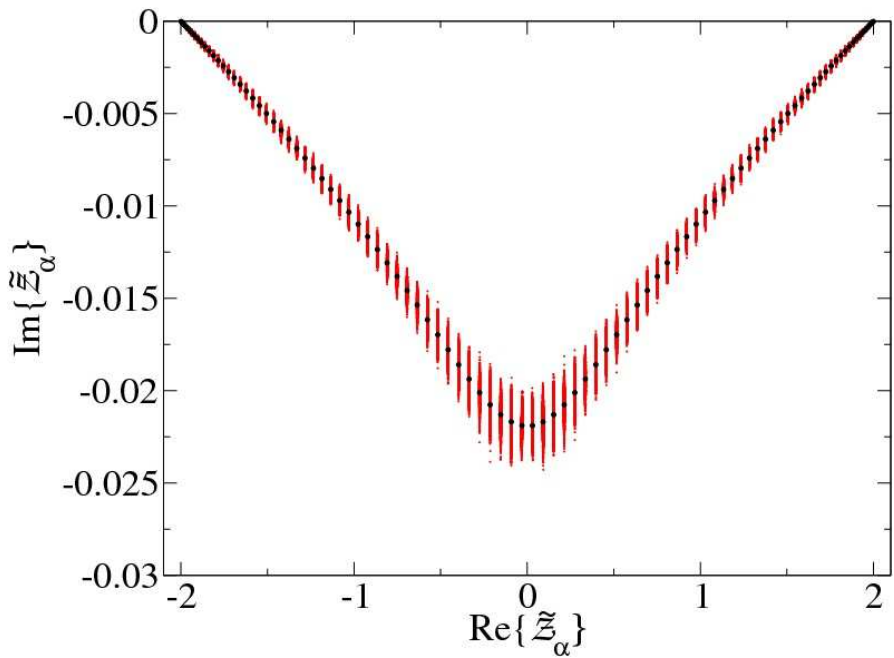}
                \caption{Scatter plot for parametric resonances where $N=100$, $\eta=0.81$ and $W=0.015$, for $5000$ realizations. As in Fig.\ref{scatter_exact}, here also dots represent the clean chain and points represent the disordered chain.}
\label{scatter_parametric}
\end{figure} 
In the next stage of the problem we switch on a very weak disorder in the chain. From a second order perturbation theory we know that for a disordered infinitely long chain the localization length, $\xi(E)$, is maximum at the middle of the band. For small $W$ it is given by \cite{krammac,thou}
\begin{eqnarray}\label{xi}
\xi(E)=\dfrac{24(4t^2-E^2)}{W^2},
\end{eqnarray}
implying thereby, $\xi(0)=\dfrac{96 t^2}{W^2}$. However, the exact result shows a small deviation at the band center due to the breakdown of the second-order perturbation theory \cite{KWegner:1981}. We consider a limiting situation when $\xi(0)/N >> 1$. For instance, in Fig. \ref{scatter_exact} and in Fig. \ref{scatter_parametric}, we show the scatter plot ($\Re\{\tilde{\mathcal{Z}}_{\alpha}\}$ vs $\Im\{\tilde{\mathcal{Z}}_{\alpha}\}$) for exact and parametric resonances respectively. In both cases we have considered $N=100$, $\eta=0.81$ and $W=0.015$ so that $\xi(0)>>N$. As seen in these figures, complex energies of the disordered chain are scattered around the $\tilde{z}_{\alpha}$s. 

We now calculate the corrections to $\tilde{z}_{\alpha}$ for such weak disorder case. It is suggestive here to deal with the self-energy. Let  $\mathcal{S}_{1}(\epsilon_{2},...,\epsilon_{N};z)$ be the self-energy for the first site, defined via
\begin{eqnarray}\label{GS}
\mathcal{G}_{11}(z)=\dfrac{1}{z-\epsilon_{1}-\mathcal{S}_{1}(\{\epsilon\};z)}.
\end{eqnarray}
Here $\{\epsilon\}$ denotes the set $\epsilon_{2},...,\epsilon_{N}$ and $\mathcal{G}_{11}$ is the $\{1,\,1\}$ element of  the resolvent $\mathcal{G}(z)=(z-\mathcal{H})^{-1}$, defined for the Hermitian matrix $\mathcal{H}$. For the later convenience we write
\begin{eqnarray}\label{Hborn}
\mathcal{H}=H+\mathcal{W}, 
\end{eqnarray}
where $\mathcal{W}=\sum_{\ell=1}^{N}\epsilon_{\ell}P_{\ell}$ and $P_{\ell}=|\,\ell\,\rangle\langle\,\ell\,|$ is the projection for the $\ell$'th site. 

In the rest of the paper we will work out results only for the exact resonances. For the parametric resonance theses results can be carried out following similar steps, so we skip all the intermediate steps merely by stating the result at the end. 

As before in Eq. (\ref{charcpol}), for disordered chain, resonances correspond to the roots of the secular equation
\begin{equation}\label{secular}
\mathcal{F}(z)=0=z-\epsilon_{1}-\mathcal{S}_{1}(\{\epsilon\};z)+\lambda\eta.
\end{equation}
Preserving $\tilde{z}_{\alpha}$ as the roots of Eq. (\ref{charcpol}), we define $\tilde{\mathcal{Z}}_{\alpha}$ as the roots of Eq. (\ref{secular}). Now we expand the roots $\tilde{\mathcal{Z}}_{\alpha}=\tilde{z}_{\alpha}+(\delta_{1} \tilde{\mathcal{Z}}_{\alpha})+(\delta_{2} \tilde{\mathcal{Z}}_{\alpha})$, assuming that $(\delta_{1}\tilde{\mathcal{Z}}_{\alpha})$ are linear while $(\delta_{2} \tilde{\mathcal{Z}}_{\alpha})$ are quadratic in the $\epsilon_{j}$ , for $j=1,...,N$. Then for $\mathcal{S}_{1}(\{\epsilon\};\tilde{\mathcal{Z}}_{\alpha})$, up to $\mathcal{O}(\{\epsilon\}^{2})$, we get 
\begin{eqnarray}\label{Selfenergy}
\mathcal{S}_{1}(\{\epsilon\};\tilde{\mathcal{Z}}_{\alpha})
&=&
S_{1}(\{0\};\tilde{z}_{\alpha})+
\sum_{n=2}^{N}\epsilon_{n}
\left(\dfrac{\partial \mathcal{S}_{1}(\{\epsilon\};z)}{\partial\epsilon_{n}}\right)_{\{\epsilon\}=0,z=\tilde{z}_{\alpha}}
\nonumber
\\
&+&
(\delta_{1} \tilde{\mathcal{Z}}_{\alpha}+\delta_{2} \tilde{\mathcal{Z}}_{\alpha})\left(\dfrac{\partial \mathcal{S}_{1}(\{\epsilon\};z)}{\partial z}\right)_{\{\epsilon\}=0,z=\tilde{z}_{\alpha}}
\nonumber
\\
&+&
\dfrac{1}{2}\sum_{n,m=2}^{N}\epsilon_{n}\epsilon_{m}
\left(\dfrac{\partial^{2} \mathcal{S}_{1}(\{\epsilon\};z)}{\partial\epsilon_{n}\partial\epsilon_{m}}\right)_{\{\epsilon\}=0,z=\tilde{z}_{\alpha}}
\nonumber
\\
&+&
\dfrac{1}{2}(\delta_{1} \tilde{\mathcal{Z}}_{\alpha})^{2}\left(\dfrac{\partial^{2} \mathcal{S}_{1}(\{\epsilon\};z)}{\partial^{2} z}\right)_{\{\epsilon\}=0,z=\tilde{z}_{\alpha}}.
\end{eqnarray}
We will use this expansion in Eq. (\ref{secular}). Before that we evaluate
\begin{eqnarray}\label{SGdz}
1-\left(\dfrac{\partial \mathcal{S}_{1}(\{\epsilon\};z)}{\partial z}\right)_{\{\epsilon\}=0,z=\tilde{z}_{\alpha}}
=
\dfrac{\partial}{\partial z} \dfrac{1}{G_{11}(z)}\bigg|_{z=\tilde{z}_{\alpha}},
\end{eqnarray}
and, 
\begin{eqnarray}\label{dGdz}
\dfrac{\partial \mathcal{S}_{1}(\{\epsilon\};z)}{\partial \epsilon_{n}}\bigg |_{\{\epsilon\}=0,z=\tilde{z}_{\alpha}}
&=&
\dfrac{1}{\left(G_{11}\right)^2}\,
\dfrac{\partial \mathcal{G}_{11}}{\partial \epsilon_{n}}\bigg|_{\{\epsilon\}=0,z=\tilde{z}_{\alpha}},
\nonumber
\\
\end{eqnarray}
for $n\geq2$. These equalities come from Eq. (\ref{GS}). Finally, we calculate derivatives of $\mathcal{G}_{11}$, at $\{\epsilon\}=0$ and $z=\tilde{z}_{\alpha}$ with respect to $\{\epsilon\}$ by using Eq. (\ref{Hborn}) for the {\it Born-series} expansion of $\mathcal{G}(z)$. We find
\begin{eqnarray}\label{dGdeps}
\dfrac{\partial \mathcal{G}_{11}}{\partial \epsilon_{n}}\bigg|_{\{\epsilon\}=0,z=\tilde{z}_{\alpha}}
&=& 
G_{1n}G_{n1}\bigg|_{z=\tilde{z}_{\alpha}}.
\end{eqnarray}
Grouping all these, for the first order corrections, we obtain 
\begin{eqnarray}\label{FPT1}
&&(\delta_{1}\tilde{\mathcal{Z}}_{\alpha})
=
\dfrac{\epsilon_{1}+\sum_{n=2}^{N}\epsilon_{n}
\dfrac{G_{1n}G_{n1}}
{\left[G_{11}\right]^{2}}\Bigg|_{z=\tilde{z}_{\alpha}}}
{\dfrac{\partial}{\partial z} \dfrac{1}{G_{11}(z)}\bigg|_{z=\tilde{z}_{\alpha}}
+
\dfrac{i\eta\exp[i\tilde{k}_{\alpha}]}
{2\sin(\tilde{k}_{\alpha})}
}. 
\end{eqnarray}
Similarly for the second order corrections we get
\begin{eqnarray}\label{dz2}
(\delta_{2} \tilde{\mathcal{Z}}_{\alpha}) 
&=&
\Bigg[\sum_{n,m=2}^{N}\epsilon_{n} \epsilon_{m}
\left\lbrace
\dfrac{G_{1n}G_{nm}G_{m1}}{[G_{11}]^{2}}
-
\dfrac{[G_{1n}G_{1m}]^{2}}
{[G_{11}]^{3}}
\right\rbrace
\nonumber
\\
&-&
\dfrac{(\delta_{1} \tilde{\mathcal{Z}}_{\alpha})^{2}}{2}
\left\lbrace\left(\dfrac{\partial^{2}}{\partial^{2} z}\dfrac{1}{G_{11}}\right)+\eta\left(\dfrac{d^{2}\exp(ik(z)}{d^{2}z}\right)\right\rbrace
\Bigg]_{\{\epsilon\}=0,z=\tilde{z}_{\alpha}}
\nonumber\\
&\times &
\Bigg[{\dfrac{\partial}{\partial z} \dfrac{1}{G_{11}(z)}\bigg|_{z=\tilde{z}_{\alpha}}
+
\dfrac{i\eta\exp[i\tilde{k}_{\alpha})}
{2\sin(\tilde{k}_{\alpha})}
}\Bigg]^{-1}. 
\nonumber\\
\end{eqnarray}

Note that $(\delta_{1} \tilde{\mathcal{Z}}_{\alpha})$ and $(\delta_{2} \tilde{\mathcal{Z}}_{\alpha}) $ have been obtained in terms of the resolvent of the closed-clean chain which we already know in terms of Chebyshev polynomials; see Eq. (\ref{G0uv}) and the relation between $u_{n}$ and $v_{n}$ with Chebyshev polynomials.


\section{Statistics of the Scattered Complex Energies}

We are interested in the statistics of the scattered complex energies. For instance, using the first order result (\ref{FPT1}) of the perturbation theory, we calculate average of square of absolute shift in complex energies defined as, $\langle|(\Delta \tilde{\mathcal{Z}}_{\alpha})|^2\rangle\equiv\langle|(\tilde{\mathcal{Z}}_{\alpha}-\tilde{z}_{\alpha})|^2\rangle$. The angular brackets are used here to represent the averaging over many realizations of set of all random site energies $\{\epsilon_{n}\}$. This quantity gives a statistical account for the scattered complex energies. We also calculate $\langle\,( \Re\{\Delta\tilde{\mathcal{Z}}_{\alpha}\})^{2}\,\rangle$ and $\langle\,(\Im\{\Delta \tilde{\mathcal{Z}}_{\alpha}\})^{2}\,\rangle$, viz, average of square of the real and the imaginary part of the shift $(\tilde{\mathcal{Z}}_{\alpha}-\tilde{z}_{\alpha})$, respectively. To obtain the latter quantities we need first to calculate $\langle\,(\Delta \tilde{\mathcal{Z}}_{\alpha})\,^{2}\rangle$ and $\langle\,[(\Delta \tilde{\mathcal{Z}}_{\alpha})^{*}]^{2}\,\rangle$, since
\begin{eqnarray}
 (\Re\{\Delta \tilde{\mathcal{Z}}_{\alpha}\})^{2}=\dfrac{(\Delta \tilde{\mathcal{Z}}_{\alpha})\,^{2}+[(\Delta \tilde{\mathcal{Z}}_{\alpha})^{*}]^{2}+2(|\Delta \tilde{\mathcal{Z}}_{\alpha})\,|^{2})}{4},
\nonumber
\\
\\
 (\Im\{\Delta\tilde{\mathcal{Z}}_{\alpha}\})^{2}=-\dfrac{(\Delta \tilde{\mathcal{Z}}_{\alpha})\,^{2}+[(\Delta \tilde{\mathcal{Z}}_{\alpha})^{*}]^{2}-2(|\Delta \tilde{\mathcal{Z}}_{\alpha})\,|^{2})}{4}.
\nonumber
\\
\end{eqnarray}
Here we have used $\{^{*}\}$ to represent the complex conjugate (c.c.).
 
For all these three statistics we simplify $(\delta_{1}\tilde{\mathcal{Z}}_{\alpha})$, given in Eq. (\ref{FPT1}), in terms of Chebyshev polynomials as
\begin{eqnarray}\label{FPT2}
(\delta_{1}\tilde{\mathcal{Z}}_{\alpha})
&=&
\Bigg[\dfrac{\tilde{z}_{\alpha}^2-4}{2}
\sum_{n=1}^{N}\epsilon_{n}
\left(U_{N-n}(\tilde{z}_{\alpha}/2)\right)^{2}
\Bigg]
\nonumber\\
&\times&
\Bigg[U_{N-1}(\tilde{z}_{\alpha}/2)T_{N+1}(\tilde{z}_{\alpha}/2)-N
\nonumber\\
&-&
i\eta\exp[i\tilde{k}_{\alpha}]\sin(\tilde{k}_{\alpha})[U_{N-1}(\tilde{z}_{\alpha}/2)]^2
\Bigg]^{-1},
\end{eqnarray}
where $T_{m}(z)=\cos[m\cos^{-1}(z)]$ is the Chebyshev polynomial of the first kind. Further simplifications occur when these polynomials are expressed in their trigonometric forms. For instance, let's calculate $|(\Delta \tilde{\mathcal{Z}}_{\alpha})|^2$, with $ \tilde{z}_{\alpha}/2=\cos( \tilde{\theta}_{\alpha})$ where $\tilde{\theta}_{\alpha}=\pi- \tilde{k}_{\alpha}$. We obtain
\begin{eqnarray}\label{abdz2}
\dfrac{|(\Delta \tilde{\mathcal{Z}}_{\alpha})|^2}{4}
=
\dfrac{\sum_{n,m=1}^{N}\epsilon_{n'}\epsilon_{m'}\sin^{2}(n' \tilde{\theta}_{\alpha})\sin^{2}(m'\tilde{\theta}^{*}_{\alpha})}
{|D(\tilde{z_{\alpha}})|^{2}}.
\end{eqnarray}
Here $n'$ and $m'$ are respectively $N+1-n$ and $N+1-m$, and $D(\tilde{z_{\alpha}})$ is simply the quantity in the second bracket of Eq. (\ref{FPT2}). Averaging releases one of the summation as the $\epsilon_{j}'$s are statistically independent-identically-distributed (i.i.d.) random variables. We simply have
\begin{eqnarray}\label{abdz3}
\langle|(\Delta \tilde{\mathcal{Z}}_{\alpha})|^2\rangle
=
\sigma^{2}\dfrac{\sum_{n=1}^{N}4\sin^{2}(n\tilde{\theta}_{\alpha})\sin^{2}(n\tilde{\theta}^{*}_{\alpha})}
{|D|^{2}},
\end{eqnarray}
where $\sigma^{2}$ is variance of the $\epsilon_{j}'$s. 

Summation in the above equality can be performed by using trigonometric identities. For instance, we first write
\begin{eqnarray}\label{abdz4}
4\sin^{2}(n\theta)\sin^{2}(n\theta^{*})
&=&
1-\cos(2n\theta)-\cos(2n\theta^{*})
\nonumber\\
&+&
\dfrac{\cos(4\,n\,\Re\{\theta\})+\cos(4\,i\,n\,\Im\{\theta\})}{2},
\end{eqnarray}
and we use the summation formula
\begin{eqnarray}\label{abdz5}
\sum_{n=1}^{N}\cos(n\theta)
&=&
\dfrac{1}{2}
\left[
\dfrac{\sin[(N+1/2)\theta]}{\sin(\theta/2)}-1
\right].
\end{eqnarray}
It turns out after some trigonometry that one can write the summation in a closed form. We find
\begin{eqnarray}
\sum_{n=1}^{N}4\sin^{2}(n\theta)\sin^{2}(n\theta^{*})
&=&N+\dfrac{1}{2}-\dfrac{U_{2N}+U_{2N}^{*}}{2}+
\dfrac{T^{*}_{2N+2}T_{2N}-T_{2N+2}T^{*}_{2N}}{2[T^{*}_{2}-T_{2}]}.
\nonumber
\\
\end{eqnarray}
Here the argument of the polynomials is $\tilde{z}_{\alpha}/2$ and for their complex conjugate it is $\tilde{z}^{*}_{\alpha}/2$. Finally, we write down finite-$N$ result for average of the absolute square of the shift, 
\begin{eqnarray}
\label{abdz6}
\langle|(\Delta \tilde{\mathcal{Z}}_{\alpha})|^2\rangle
=
\sigma^{2}\dfrac{N+\dfrac{1}{2}-\dfrac{U_{2N}+U_{2N}^{*}}{2}
+
\dfrac{T^{*}_{2N+2}T_{2N}-T_{2N+2}T^{*}_{2N}}{2[T^{*}_{2}-T_{2}]}}
{|D|^{2}}.
\end{eqnarray}

We now turn our attention to large-$N$ behavior of the result (\ref{abdz6}). For this purpose we use the ansatz (\ref{ansatz}) and result the (\ref{res}) for $\tilde{k}_{\alpha}$. Large-N behavior for the Chebyshev polynomials, with argument $\tilde{z}_{\alpha}$, may be calculated as
\begin{eqnarray}
T_{2N}(\tilde{z}_{\alpha}/2)&=&\dfrac{\exp[2iN\tilde{\theta}_{\alpha}]+\exp[-2iN\tilde{\theta}_{\alpha}]}{2} 
\nonumber
\\
&\approx&
\dfrac{\Omega({k}_{\alpha};\eta)\exp(-2i{k}_{\alpha})+\left[\Omega({k}_{\alpha};\eta)\right]^{-1}\exp(2i{k}_{\alpha})}{2},
\end{eqnarray}
\begin{eqnarray}
T_{2N+2}(\tilde{z}_{\alpha}/2)
=
\dfrac{\Omega(k_{\alpha};\eta)+\left[\Omega(k_{\alpha};\eta)\right]^{-1}}{2}+\mathcal{O}(N^{-1}),
\end{eqnarray}
\begin{eqnarray}
U_{2N}(\tilde{z}_{\alpha}/2)&=&\dfrac{\exp[i(2N+1)\tilde{\theta}_{\alpha}]-\exp[-i(2N+1)\tilde{\theta}_{\alpha}]}
{\exp(i\tilde{\theta}_{\alpha})-\exp(-i\tilde{\theta}_{\alpha})} 
\nonumber
\\
&\approx&
\dfrac{\Omega({k}_{\alpha};\eta)\exp(-i{k}_{\alpha})-\left[\Omega({k}_{\alpha};\eta)\right]^{-1}\exp(i{k}_{\alpha})}{\exp(-i{k}_{\alpha})-\exp(i{k}_{\alpha})}.
\end{eqnarray}
Finally, 
\begin{eqnarray}
 T_{2}(\tilde{z}^{*}_{\alpha}/2)- T_{2}(\tilde{z}_{\alpha}/2)
&=& -\dfrac{2i\,\Im\{\Phi_{\alpha}\}}{N}\,
{z}_{\alpha}+\mathcal{O}(N^{-2})
\nonumber
\\
&\approx&
\dfrac{4i}{N}\, \cos(k_{\alpha})\,\Im\{\Phi_{\alpha}\},
\end{eqnarray}
where we have used the ansatz (\ref{ansatz}) in the second order polynomial $T_{2}(z)=2z^{2}-1$ and $\Im\{\Phi_{\alpha}\}=-\sin({k}_{\alpha})\,\text{ln}(|\Omega|)$, as obtained from Eqs. (\ref{ansatz}, \ref{res}). 

We can now plug in these results in Eq. (\ref{abdz6}). These asymptotic results gives the numerator as ($N+a1+a2/(a3/N)$) where $a1,\,a2/a3$ are $\mathcal{O}(N^{0})$. Similarly we obtain denominator as ($\,N^{2}+b1\,N+b2$) where $b1$ and $b2$ are $\mathcal{O}(N^{0})$; see Appendix A for details. Thus in the leading order, we obtain
\begin{eqnarray}\label{abdz7}
\langle|(\Delta \tilde{\mathcal{Z}}_{\alpha})|^2\rangle
&=&
\dfrac{\sigma^{2}}{N}\,
\left(
1+\dfrac{1}{8}\,
\dfrac{
\left(
|\Omega|^{2}-|\Omega|^{-2}
\right)}
{\text{ln}(|\Omega|)}
\right).
\end{eqnarray}

\begin{figure}
        \centering
               \includegraphics [width=0.75\textwidth]{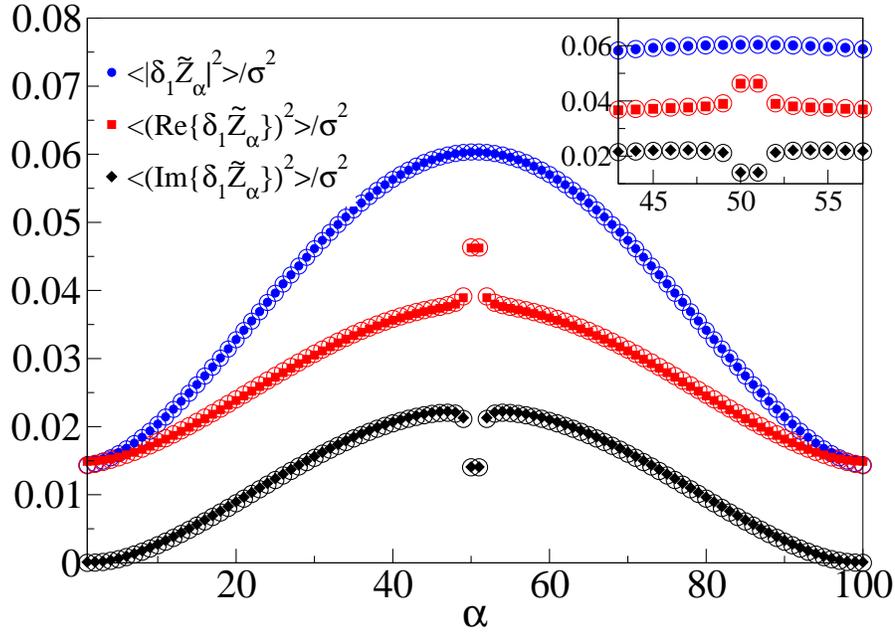}
                \caption{Asymptotic results for $\langle\,|\delta_{1}\tilde{\mathcal{Z}}_{\alpha}|^{2}\,\rangle/\sigma^{2}$, $\langle\,(\Re\{\delta_{1}\tilde{\mathcal{Z}}_{\alpha}\})^{2}\,\rangle/\sigma^{2}$ and $\langle\,(\Im\{\delta_{1}\tilde{\mathcal{Z}}_{\alpha}\})^{2}\,\rangle/\sigma^{2}$, shown by filled circles, squares and diamonds, vs energy index $\alpha$. We have compared here the finite-$N$ results, shown by open circles, for the exact resonances where $N=100$ and $\eta=0.81$. In the set we show these results for 14 energy indices near the middle of the energy band but on a different scale.}
\label{ex_finitevsLarge}
\end{figure} 

What follows next is the calculation of large-$N$ results for $\langle\,( \Re\{\Delta\tilde{\mathcal{Z}}_{\alpha}\})^{2}\,\rangle$ and $\langle\,(\Im\{\Delta \tilde{\mathcal{Z}}_{\alpha}\})^{2}\,\rangle$. Since we need first to calculate $\langle\,(\Delta \tilde{\mathcal{Z}}_{\alpha})\,^{2}\rangle$ and $\langle\,[(\Delta \tilde{\mathcal{Z}}_{\alpha})^{*}]^{2}\,\rangle$, from Eq. (\ref{FPT2}) we obtain after averaging
\begin{eqnarray}\label{dzsq}
 \langle\,(\Delta \tilde{\mathcal{Z}}_{\alpha})^{2}\,\rangle
=\sigma^{2}\dfrac{\sum_{n=1}^{N}4\sin^{4}(n\tilde{\theta}_{\alpha})}
{D^{2}},
\end{eqnarray}
and
\begin{eqnarray}\label{cdzsq}
 \langle\,[(\Delta \tilde{\mathcal{Z}}_{\alpha})^{*}]^{2}\,\rangle
=\sigma^{2}\dfrac{\sum_{n=1}^{N}4\sin^{4}(n\tilde{\theta}^{*}_{\alpha})}
{(D^{*})^{2}}.
\end{eqnarray}
\begin{figure}
        \centering
               \includegraphics [width=0.75\textwidth]{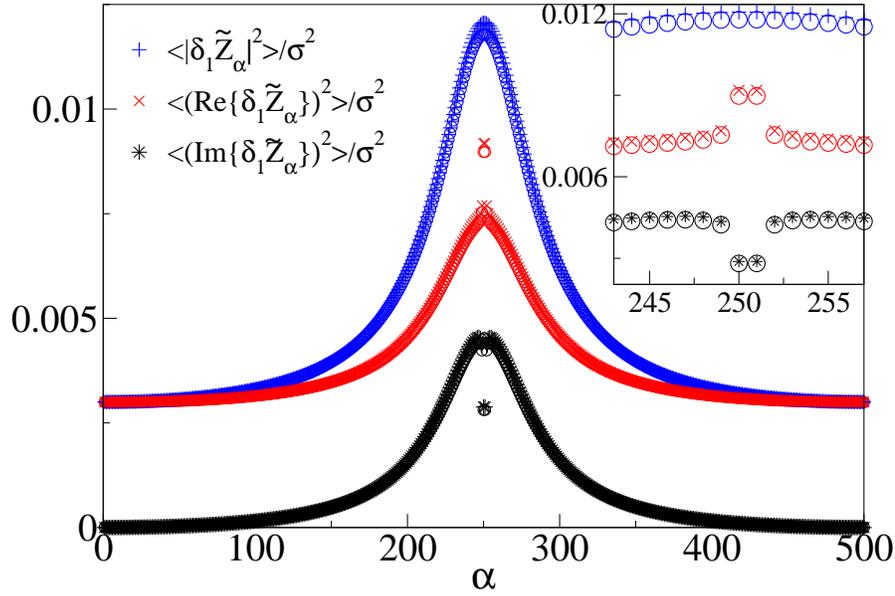}
                \caption{Shown on the same pattern of Fig. \ref{ex_finitevsLarge} but for the parametric resonances where $N=500$ and $\eta=0.81$. In this figure $\langle\,|\delta_{1}\tilde{\mathcal{Z}}_{\alpha}|^{2}\,\rangle/\sigma^{2}$, $\langle\,(\Re\{\delta_{1}\tilde{\mathcal{Z}}_{\alpha}\})^{2}\,\rangle/\sigma^{2}$ and $\langle\,(\Im\{\delta_{1}\tilde{\mathcal{Z}}_{\alpha}\})^{2}\,\rangle/\sigma^{2}$ are shown respectively by pluses, crosses and stars. The  inset is shown for the indices near the middle of the band.}
\label{pa_finitevsLarge}
\end{figure} 
For summation we use the formula \cite{GR}
\begin{eqnarray}\label{sumsin4}
\sum_{n=1}^{N} \sin^{4}(n\,\theta)
&=&
\dfrac{1}{8}
\Bigg[
3N-\dfrac{\sin(N\theta)}{\sin(\theta)}
\big(4\cos[(N+1)\theta]
\nonumber
\\
&-&
\dfrac{\cos[2(N+1)\theta]\,\cos(N\theta)}{\cos(\theta)}
\big)\,
\Bigg],
\nonumber
\\
\sum_{n=1}^{N} \sin^{4}(n\,\tilde{\theta}_{\alpha})&=&
\dfrac{1}{8}
\left[
3N-4U_{N-1}T_{N+1}+\dfrac{T_{2N+2}U_{2N-1}}{\tilde{z}_{\alpha}}
\right],
\end{eqnarray}
where in the second equality the polynomials have argument $\tilde{z}_{\alpha}/2$ with $2\cos(\tilde{\theta}_{\alpha})=\tilde{z}_{\alpha}$. Similarly for the summation in Eq. (\ref{cdzsq}) one gets the polynomials with argument $\tilde{z}^{*}_{\alpha}/2$. One can now use the equality (\ref{sumsin4}) in Eqs. (\ref{cdzsq}, \ref{cdzsq}) in order to derive finite-$N$ result for $\langle\,(\Re\{\Delta \tilde{\mathcal{Z}}_{\alpha}\})^{2}\,\rangle$ and $\langle\,(\Im\{\Delta \tilde{\mathcal{Z}}_{\alpha}\})^{2}\,\rangle$. For large-$N$ we make use of the ansatz (\ref{ansatz}) and calculate the leading order contribution as
\begin{eqnarray}\label{LargeNdx}
\langle\,(\Re\{\Delta \tilde{\mathcal{Z}}_{\alpha}\})^{2}\,\rangle
&=& 
\dfrac{\sigma^{2}}{2N}
\Bigg\{\dfrac{5}{2}+\dfrac{1}{8}\,
\dfrac{
\left(
|\Omega|^{2}-|\Omega|^{-2}
\right)}
{\text{ln}(|\Omega|)}
+
g(k_{\alpha})\Bigg\},
\\
\nonumber
\\
\label{LargeNdy}
\langle\,(\Im\{\Delta \tilde{\mathcal{Z}}_{\alpha}\})^{2}\,\rangle
&=&
-\dfrac{\sigma^{2}}{2N}
\Bigg\{\dfrac{1}{2}-\dfrac{1}{8}\,
\dfrac{
\left(
|\Omega|^{2}-|\Omega|^{-2}
\right)}
{\text{ln}(|\Omega|)}
+
g(k_{\alpha})\Bigg\},
\end{eqnarray}
where
\begin{eqnarray}
g(k_{\alpha})&=&
\dfrac{1}{8}
\left(
\dfrac{e^{-2ik_{\alpha}}\Omega^{2}-e^{2ik_{\alpha}}\Omega^{-2}}
{4i \sin(k_{\alpha})\left(2N\cos(k_{\alpha})+i\sin(k_{\alpha})\text{ln}(\Omega)\right)}
\right) 
+
(\text{c.c.}).
\end{eqnarray}

Equations (\ref{abdz7}, \ref{LargeNdx}, \ref{LargeNdy}) are our main analytical results and they are also valid for parametric resonances with the $\Omega$ given in Eq. (\ref{Omgpara}). In Fig. \ref{ex_finitevsLarge} we verify the asymptotic results (\ref{abdz7}, \ref{LargeNdx}, \ref{LargeNdy}) against their finite-$N$ counterparts, for exact resonances with $N=100$ and $\eta=0.81$. Fig. \ref{pa_finitevsLarge} is repeated on the same pattern but for parametric resonances where $N=500$ and $\eta=0.81$.  They confirm that the asymptotic results give a good account for the finite-$N$ results. However, there are some exception near the edges (not visible on the scale of the plot) where the ansatz (\ref{ansatz}) is not valid. 

It turns out that in order to calculate the DOR we need the second order corrections $(\delta_{2}\tilde{\mathcal{Z}}_{\alpha})$, derived in Eq. (\ref{dz2}). We have followed the method used earlier \cite{FZ:99} for Hatano-Nelson Model \cite{HN:97}. However, we have not been able to obtain a closed expression of the DOR. This is discussed in Appendix B where we leave the calculations with a formal expression for the DOR. 


\section{Numerical Methods and Verification of The Eqs. (\ref{abdz7}, \ref{LargeNdx}, \ref{LargeNdy})}
\begin{figure}
        \centering
               \includegraphics [width=0.75\textwidth]{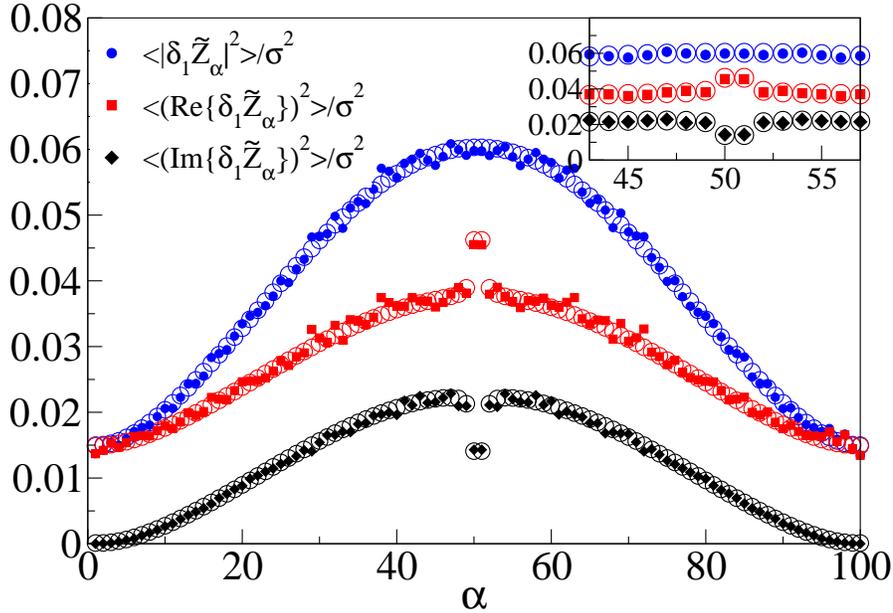}
                \caption{Comparison of asymptotic results with numerics, for exact resonances where $N=100$, $W=0.015$ and $\eta=0.81$. In this figure, filled circles, squares and diamonds are the numerical results respectively for $\langle\,|\delta_{1}\tilde{\mathcal{Z}}_{\alpha}|^{2}\,\rangle/\sigma^{2}$, $\langle\,(\Re\{\delta_{1}\tilde{\mathcal{Z}}_{\alpha}\})^{2}\,\rangle/\sigma^{2}$ and $\langle\,(\Im\{\delta_{1}\tilde{\mathcal{Z}}_{\alpha}\})^{2}\,\rangle/\sigma^{2}$ while open circles are the rescaled theories (\ref{abdz7}, \ref{LargeNdx}) and (\ref{LargeNdy}). In the inset we show a comparison for 14 indices near the middle of the energy band on a different scale of the plot.}
\label{ex_numericsvsLarge}
\end{figure}

Numerical simulations for parametric resonance are always cost efficient. The reason being that there one deals with standard eigenvalue problem for which many fast subroutine packages are available, for instance LAPACK. On the other hand to verify the results (\ref{abdz7}, \ref{LargeNdx}, \ref{LargeNdy}) for exact resonances, where one needs to obtain numerical solutions of a characteristic polynomial equation of order $N$ in a complex plane, there is no as good algorithm. In this paper we show results for the exact resonances by calculating roots of the characteristic polynomial where we have used a cost efficient numerical subroutine {\it ezero}. The subroutine is available on the CPC program library. There is one major advantage of using this subroutine over other methods, for instance the Newton's method. This subroutine does not require initial guesses for the roots but only the contour which encloses all the roots of the polynomial. Besides, it also avoids calculating the derivatives which may result into numerical overflow. 

In alternative to {\it ezero} we have used a different approach for calculating the roots. We survey the complex $\tilde{k}$-plane for the zeros of the $\text{Det}[M(\tilde{k})M(\tilde{k})^{\dagger}]$ where $M_{rs}=-2\cos(\tilde{k})\delta_{rs}-\tilde{\mathcal{H}}_{rs}$ for $r,s=1,...,N$ \cite{Neuberger}. (In our system $-\pi<\Re\{\tilde{k}\}<\pi$ and $\Im\{\tilde{k}\}<0$.) These zeros give the eigenvalues of $\tilde{\mathcal{H}}$. However, in the latter approach it is advisable to disintegrate the complex plane into small cells at first and then at every iteration into smaller one - only for $N$ cells which contain minima of the lowest eigenvalue and throwing the rest out. In this way one makes the algorithm faster and obtain the zeros in a reasonable precision. For a tridiagonal matrix this algorithm consumes a time which roughly grows with $N^3$. However, while comparing the two methods on a simple machine we find that the method used in {\it ezero} is much faster than the method described here. We refer to \cite{ezero} for further details of this subroutine.  

\begin{figure}
        \centering
               \includegraphics [width=0.75\textwidth]{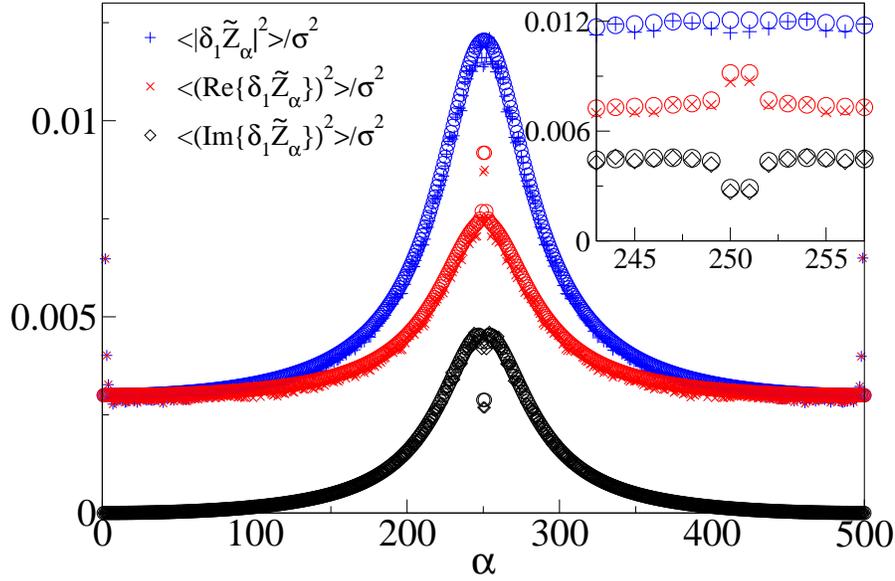}
                \caption{ Shown on the same pattern of Fig. \ref{ex_numericsvsLarge} but for the parametric resonances where $N=500$, $W=0.015$ and $\eta=0.81$. These numerical results are obtained from the diagonalization of $N$-dimensional matrices for  $5000$ realizations. In this figure $\langle\,|\delta_{1}\tilde{\mathcal{Z}}_{\alpha}|^{2}\,\rangle/\sigma^{2}$, $\langle\,(\Re\{\delta_{1}\tilde{\mathcal{Z}}_{\alpha}\})^{2}\,\rangle/\sigma^{2}$ and $\langle\,(\Im\{\delta_{1}\tilde{\mathcal{Z}}_{\alpha}\})^{2}\,\rangle/\sigma^{2}$ are shown respectively by pluses, crosses and stars while open circles are the rescaled theories (\ref{abdz7}, \ref{LargeNdx}) and (\ref{LargeNdy}).}
\label{pa_numericsvsLarge}
\end{figure}

In Fig. \ref{ex_numericsvsLarge}, we compare asymptotic results with simulation done for the total number of realizations $L=2500$, for exact resonances. In Fig. \ref{pa_numericsvsLarge} we compare numerical results obtained for parametric resonances, where $N=500$, $\eta=0.81$ and $L=5000$, with our theory for large-$N$. Though we have considered only the flat disorder yet our results are valid for the Gaussian or other symmetric distribution functions. These figures show that our asymptotic results are in fair agreement with the numerical results for almost all $\alpha$. For instance, near the middle of the energy band it describes reasonably well a dip and a peak, respectively in the $\langle\,(\Im\{\Delta \tilde{\mathcal{Z}}_{\alpha}\})^{2}\,\rangle$ and $\langle\,(\Re\{\Delta \tilde{\mathcal{Z}}_{\alpha}\})^{2}\,\rangle$. These two opposite effects, however, cancel out in  $\langle\,(|\{\Delta \tilde{\mathcal{Z}}_{\alpha}\}|)^{2}\,\rangle$.

\section{Conclusion}

In conclusion, we have studied resonances in a one dimensional discrete tight-binding open chain in a weak disorder limit. In this study we have calculated complex energies in an open-clean chain of finite length. The result we obtain is a polynomial equation which we have been able to solve for long chains using an ansatz for the solution. To the best of our knowledge, this result has never been derived before. We have used a perturbation theory up to the second order where we have derived the first and the second order corrections to the complex energies in terms of Chebyshev polynomials. The first order corrections have been useful to obtain closed form of the statistical results for the scattered complex energies. These results have been further simplified for long chains where we obtain compact results. The asymptotic results have been verified against numerics. Evidently, in the weak disorder limit the perturbation theory predicts nice statistical results. Our results are new in these studies and they could be useful in the further studies of such systems. 

It would be interesting to study statistics of resonances in the weak disorder limit for higher dimensional models as well as for the cases when the site energies are not independent random variables but they are correlated with each other \cite{Izrailev:99}. Besides, there has been growing interest for the case when $M$ sites are connected to the outer world where $1\le M\le N$ \cite{Borgonovi:2012}. We believe that our methods could be useful for the study of such models. Finally, we mention the case where  $\xi(0)\sim\mathcal{O}(N)$. It requires a separate investigation as  our perturbative analysis fails in this limit.

\ack
The author is thankful to Boris Shapiro for suggesting the problem to him. The author would also like to give credit to Joshua Feinberg for the derivation of some of the equations in Secs. III and IV and also for the help in Appendix B. Discussions with both of them are gratefully acknowledged. The author also acknowledges Marko \v Znidari\v c and Thomas H. Seligman for reading the manuscript. 

Support from ISF-1067 and generous hospitality of Technion Institute are also acknowledged. Additional support by the project 79613 by CONACyT, Mexico, is acknowledged.

\appendix
\section{Large-$N$ behavior of the denominator in (\ref{FPT2})}
The denominator in Eq. (\ref{FPT2}) can be simplified as follows:
\begin{eqnarray}\label{app1}
D(z_{\alpha})
&=&
-N+U_{N-1}(z_{\alpha}/2)T_{N+1}(z_{\alpha}/2)-i\eta\exp[I(z_{\alpha})]
\nonumber
\\
&\times&
\sin[k(z_{\alpha})][U_{N-1}(z_{\alpha}/2)]^{2}
\nonumber
\\
&\approx&
-N+\Big[-\exp(-ik_{\alpha})[\Omega(1-\eta)+1]
\nonumber
\\
&+&
\exp(ik_{\alpha})[\Omega^{-1}(1+\eta\exp(2ik_{\alpha}))+1-2\eta]
\Big]
\nonumber
\\
&\times&
\Big[2[\exp(-ik_{\alpha})-\exp(ik_{\alpha})]\Big]^{-1}.
\end{eqnarray}
Using now
\begin{equation}
\Omega(1-\eta)+1=2-\eta \exp(2ik_{\alpha}),
\end{equation}
and
\begin{equation}
\Omega^{-1}(1+\eta\exp(2ik_{\alpha}))+1-2\eta
=
\dfrac{2-3\eta+\eta^{2}\exp(2ik_{\alpha})}
{1-\eta\exp(2ik_{\alpha})},
\nonumber
\\
\end{equation}
in (\ref{app1}) we get
\begin{eqnarray}
D
&\simeq& 
-N-\dfrac{1}{1-\eta\exp(2ik_{\alpha})}.
\end{eqnarray}
For $|D|^{2}$ this yields
\begin{eqnarray}
 |D|^{2}
&=&
N^{2}+N\,\dfrac{1-\eta\cos(2k_{\alpha})}{1+\eta^{2}-2\eta\cos(2k_{\alpha})}
\nonumber
\\
&+&
\dfrac{1}{1+\eta^{2}-2\eta\cos(2k_{\alpha})}.
\end{eqnarray}

Similarly for the parametric resonances, the denominator for large $N$ is given by
\begin{eqnarray}
D
&=&
-N+U_{N-1}(z_{\alpha}/2)T_{N+1}(z_{\alpha}/2)
\nonumber
\\
&\simeq&
-(N+1/2)+
\dfrac{\exp(-ik_{\alpha})\Omega-\exp(ik_{\alpha})\Omega^{-1}}{\exp(ik_{\alpha})-\exp(-ik_{\alpha})} 
\nonumber
\\
&\simeq&
-(N+1/2)
+\dfrac{(1+\eta^{2})\,\Omega}{2[\eta\exp(ik_{\alpha})+i]^{2}}.
\end{eqnarray}
Thus for $|D|^{2}$ we get
\begin{eqnarray}
|D|^{2}&=&N^{2} +N\,\dfrac{(1-\eta^{4})}{(1+\eta^{2})^{2}-4\eta^{2}\sin^{2}(k_{\alpha})}
\nonumber
\\
&+&
\dfrac{(1+\eta^{2})^{2}}{4[1+\eta^{2})^{2}-4\eta^{2}\sin^{2}(k_{\alpha})]}
\,
\dfrac{1+\eta^{2}+2\eta\sin(k_{\alpha})}
{1+\eta^{2}-2\eta\sin(k_{\alpha})}.
\nonumber
\\
\end{eqnarray}
Clearly in both cases $|D|^{2}$ has a form $N^{2}+b1\, \{N^{1}\}+b2\,\{N^{0}\}$.


\section{Calculation of DOR}

We define the average DOR as
\begin{eqnarray}\label{DOS1}
\langle\,\rho(x,y)\,\rangle=\left\langle\,\sum_{\alpha=1}^{N}\delta(x-\mathcal{\tilde{X}}_{\alpha})\delta(y-\mathcal{\tilde{Y}}_{\alpha})\,\right\rangle,
\end{eqnarray}
where $\mathcal{\tilde{X}}_{\alpha}\equiv\Re\{\tilde{\mathcal{Z}}_{\alpha}\}$ and $\mathcal{\tilde{Y}}_{\alpha}\equiv\Im\{\tilde{\mathcal{Z}}_{\alpha}\}$. Next we define  $\tilde{x}_{\alpha}=\Re\{\tilde{z}_{\alpha}\}$, $\tilde{y}_{\alpha}=\Im\{\tilde{z}_{\alpha}\}$, $(\delta_{1}\tilde{x}_{\alpha})=\Re\{(\delta_{1}\tilde{\mathcal{Z}}_{\alpha})\}$, $(\delta_{1}\tilde{y}_{\alpha})=\Im\{(\delta_{1}\tilde{\mathcal{Z}}_{\alpha})\}$, $(\delta_{2}\tilde{x}_{\alpha})=\Re\{(\delta_{2}\tilde{\mathcal{Z}}_{\alpha})\}$ and 
$(\delta_{2}\tilde{y}_{\alpha})=\Im\{(\delta_{2}\tilde{\mathcal{Z}}_{\alpha})\}$ and then use the expansion
$\mathcal{\tilde{X}}_{\alpha}=\tilde{x}_{\alpha}+(\delta_{1}\tilde{x}_{\alpha})+(\delta_{2}\tilde{x}_{\alpha})$ and $\mathcal{\tilde{Y}}_{\alpha}=\tilde{y}_{\alpha}+(\delta_{1}\tilde{y}_{\alpha})+(\delta_{2}\tilde{y}_{\alpha})$ in (\ref{DOS1}). We find
\begin{eqnarray}\label{DOS2}
&&\langle\,\rho(x,y)\,\rangle
=
\sum_{\alpha=1}^{N}\delta(x-\tilde{x}_{\alpha})\delta(y-\tilde{y}_{\alpha})
\nonumber\\
&+&
\sum_{\alpha=1}^{N}
\Big[
\langle
(\delta_{1}\tilde{x}_{\alpha})(\delta_{1}\tilde{y}_{\alpha}) 
\rangle
\delta'(x-\tilde{x}_{\alpha})\delta'(y-\tilde{y}_{\alpha})
-
\langle
(\delta_{2}\tilde{x}_{\alpha})
\rangle
\delta'(x-\tilde{x}_{\alpha})\delta(y-\tilde{y}_{\alpha})
\nonumber
\\
&-&
\langle
(\delta_{2}\tilde{y}_{\alpha})
\rangle
\delta(x-\tilde{x}_{\alpha})\delta'(y-\tilde{y}_{\alpha})
+
\dfrac{1}{2}
\langle(\delta_{1}\tilde{x}_{\alpha})^{2}
\rangle
\delta''(x-\tilde{x}_{\alpha})\delta(y-\tilde{y}_{\alpha})
\nonumber\\
&+&
\dfrac{1}{2}
\langle(\delta_{1}\tilde{y}_{\alpha})^{2}
\rangle
\delta(x-\tilde{x}_{\alpha})\delta''(y-\tilde{y}_{\alpha})
\Big]. 
\end{eqnarray}
Here $\delta'(x)=d\,\delta(x)/dx$ and similarly $\delta''(x)$ is the second derivative of the Dirac-Delta function with respect to the argument. We have already shown that $\langle(\delta_{1}\tilde{x}_{\alpha})^{2}\rangle$ and $\langle(\delta_{1}\tilde{y}_{\alpha})^{2}\rangle$ are of $\mathcal{O}(\sigma^{2}N^{-1})$ while $\langle(\delta_{1}\tilde{x}_{\alpha})(\delta_{1}\tilde{y}_{\alpha})\rangle$ is also $\mathcal{O}(\sigma^{2}N^{-1})$ since 
\begin{eqnarray}
\langle(\delta_{1}\tilde{x}_{\alpha})(\delta_{1}\tilde{y}_{\alpha})\rangle
=
\dfrac{\langle(\delta_{1}\mathcal{\tilde{Z}}_{\alpha})^{2}\rangle-\langle([\delta_{1}\mathcal{\tilde{Z}}_{\alpha}]^{*})^{2}\rangle}
{4i}. 
\end{eqnarray}

Motivated from \cite{FZ:99}, we calculate the coefficient of $\epsilon_{n}^{2}$ in (\ref{dz2}) to obtain $\langle(\delta_{2}\tilde{x}_{\alpha})\rangle$ and $\langle(\delta_{2}\tilde{y}_{\alpha})\rangle$. Let's write
\begin{eqnarray}\label{cn1}
\delta_{1}\tilde{\mathcal{Z}}_{\alpha}
=
\sum_{n=1}^{N}\epsilon_{n}c_{n;\alpha}, 
\end{eqnarray}
and
\begin{eqnarray}\label{cn2}
\delta_{2}\tilde{\mathcal{Z}}_{\alpha} 
=
\sum_{n,m=1}^{N}\epsilon_{n}\epsilon_{m}d_{nm;\alpha}.
\end{eqnarray}
Expressing the polynomials in (\ref{FPT2}) in terms of sinusoidal functions, we simply read-off $c_{n;\alpha}$:
\begin{eqnarray}\label{cn3}
c_{n;\alpha}&=&
\dfrac{-2\sin^{2}(n'\theta_{\alpha})}{D_{\alpha}} 
\nonumber
\\
&=&
\dfrac{\cos(2n'\theta_{\alpha})-1}{D_{\alpha}}.
\end{eqnarray}
Here $n'=N+1-n$ and $D_{\alpha}$ is the denominator inside the brackets of (\ref{FPT2}). The denominator is $\mathcal{O}(N)$ thus $c_{n;\alpha}\sim \mathcal{O}(N^{-2})$, hence will be dropped off. Using now our ansatz, for $d_{nn;\alpha}$ we obtain
\begin{eqnarray}\label{dnn2}
d_{nn;\alpha}
&=&
\dfrac{(1-\cos(2r\Psi_{\alpha}))}{4\,N\,\sin(\Psi_{\alpha}/N)\sin(\Psi_{\alpha})}
\Big\{
4\cos(2 r \Psi_{\alpha})
\nonumber
\\
&-&
\cos[(1+2r)\Psi_{\alpha}]-3\cos[(1-2r)\Psi_{\alpha}]
\Big\}, 
\end{eqnarray}
where $r=n'/N$ and $\Psi_{\alpha}=Nk_{\alpha}$ with $k_{\alpha}$ given in (\ref{res}). There is no further simplification of this result to obtain a compact and simple expression for real and imaginary parts of $d_{nn;\alpha}$ (as the authors \cite{FZ:99} have been able to do for with the result they obtain for the Hatano-Nelson model \cite{HN:97}). So we leave the density formally as
\begin{eqnarray}
&&\langle\rho(x,y)\rangle 
=
\rho_{0}(x,y)-\sigma^{2}\sum_{\alpha=1}^{N}
\Big(\delta'(x-x_{\alpha})\delta(y-y_{\alpha})
\nonumber
\\
&\times&
\sum_{n=1}^{N}\Re\{d_{nn;\alpha}\}
+
\delta(x-x_{\alpha})\delta'(y-y_{\alpha})
\sum_{n=1}^{N}\Im\{d_{nn;\alpha}\}
\Big),
\end{eqnarray}
where $\rho_{0}(x,y)$ is first term of Eq. (\ref{DOS2}).



\begin{thebibliography}{}
\bibitem{Neuberger}  The method is suggested by H.~Neuberger and then developed and tested by the author on the system studied herein.
\end{thebibliography}
\end{document}